\RequirePackage{ifpdf}
\ifpdf 
\documentclass[pdftex]{sigma}
\else
\documentclass{sigma}
\fi

\begin{document}

\renewcommand{\PaperNumber}{017}

\FirstPageHeading

\ShortArticleName{Subgroups of the Group of Generalized Lorentz
Transformations}

\ArticleName{Subgroups of the Group of Generalized Lorentz\\
Transformations and Their Geometric Invariants}

\Author{George Yu. BOGOSLOVSKY}

\AuthorNameForHeading{G.Yu. Bogoslovsky}

\Address{Skobeltsyn Institute of Nuclear Physics, Moscow State
University, 119992 Moscow, Russia}

\Email{\href{mailto:bogoslov@theory.sinp.msu.ru}{bogoslov@theory.sinp.msu.ru}}

\ArticleDates{Received October 06, 2005, in final form November
09, 2005; Published online November 15, 2005}

\Abstract{It is shown that the group of generalized Lorentz
transformations serves as relativistic symmetry group of a flat
Finslerian event space. Being the generalization of Minkowski
space, the Finslerian event space arises from the spontaneous
breaking of initial gauge symmetry and from the formation of
anisotropic fermion-antifermion condensate. The principle of
generalized Lorentz invariance enables exact taking into account
the influence of  condensate on the dynamics of fundamental
fields. In particular, the corresponding genera\-lized Dirac
equation turns out to be nonlinear. We have found two noncompact
subgroups of the group of generalized Lorentz symmetry and their
geometric invariants. These subgroups play a key role in
constructing  exact solutions of such equation.}


\Classification{53C60; 53C80}

\vspace{-2mm}

\section{Introduction}

It is common knowledge that since its discovery and for a hundred
years ahead the Lorentz symmetry has determined the development of
the theory of fundamental interactions. At present, however, owing
to progress made in the construction of unified gauge theories,
there have been grounds to consider the Lorentz symmetry as not
strict but only approximate symmetry of nature.

According to the most popular point of view, even in a theory
which has Lorentz invariance at the fundamental level, this
symmetry can be spontaneously broken if some (for example, vector)
field acquires a vacuum expectation value which breaks the initial
Lorentz symmetry. Certainly the question here concerns symmetry
violation with respect to active Lorentz transformations of
fundamental fields against the background of fixed (in this case,
vector) condensate. As to the passive Lorentz transformations
under which the condensate is transformed as a Lorentz vector, the
corresponding Lorentz covariance remains valid.

Noteworthy is the fact that the usual Standard Model of strong,
weak and electromagnetic interactions does not have the dynamics
necessary to  cause spontaneous breaking of the Lorentz symmetry.
In other words, the standard Higgs mechanism, which breaks the
local gauge symmetry and gives rise to the scalar Higgs
condensate, does not affect the initial Lorentz symmetry of the
theory.

In order to describe possible effects caused by violation of
active Lorentz invariance and to classify them as effects of
Planck-scale physics, strongly suppressed at attainable energy
scales, the so-called Standard Model Extension has been proposed
\cite{kostelecky&samuel,colladay&kostelecky,kostelecky2004,kosteleckyEd}.
The Lagrangian of this phenomenological theory is constructed so
that it includes all passive Lorentz scalars formed by combining
standard-model fields with coupling coefficients having Lorentz
indices. As a result, along with the Lorentz symmetry the
relativistic symmetry turns out to be ad-hoc violated.

In contrast to such string-motivated theory (see also
\cite{allen&yokoo,cardone&mignani})
 there exists another, Finslerian approach to the problem
 \cite{bogoslovsky1977,bogoslovskyBook,bogoslovsky19929394,bogoslovsky&goenner199899},
 which permits Lorentz symmetry violation without violation of relativistic symmetry.
 It is based on the following idea. Spontaneous breaking of the initial gauge symmetry
 may be accompanied by the corresponding phase transition in
 the geometrical structure of space-time. In other words, spontaneous
 breaking of the gauge symmetry may lead to a dynamic rearrangement of
 the vacuum which results in the formation of a relativistically invariant
 anisotropic fermion-antifermion condensate, i.e.\ of a constant
 classical nonscalar field\footnote{It is worth noting that,
irrespective of the problem of violation of Lorentz symmetry, in
the literature consideration has already been given to the
mechanism of the dynamical breaking of the initial gauge symmetry
which is alternative to the standard one; instead of the
elementary Higgs condensate there appears a scalar
fermion-antifermion condensate \cite{arbuzov}.}. This constant
field physically manifests itself as a relativistically invariant
anisotropic medium filling space-time. Such a medium, leaving
space-time flat, gives rise to its anisotropy, that is, instead of
Minkowski space, there appears a relativistically invariant
Finslerian event space with partially broken 3D isotropy.

\section{The group of isometries of the flat Finslerian space-time\\ with partially broken 3D isotropy}

In comparison with the 6-parameter homogeneous Lorentz group of
Minkowski space, the homo\-geneous group of isometries of the flat
Finslerian space with partially broken 3D isotropy is 
a~4-parameter group. Apart from 3-parameter boosts (generalized
Lorentz transformations) it includes only axial symmetry
transformations, i.e.\ the 1-parameter group of rotations around
the preferred direction in 3D space, in which case this direction
is determined by a spontaneously arising axially symmetric
fermion-antifermion condensate. It will be demonstrated below that
the 4-parameter group of Finslerian isometries is locally
isomorphic to the corresponding \mbox{4-pa\-rameter} subgroup of the
Lorentz group. Therefore we primarily consider the 4-parameter
subgroup of the Lorentz group.

\subsection{The 4-parameter subgroup of Lorentz group\\ and its 3-parameter noncompact subgroup}

In \cite{wintwrnitz&fris}, in terms of Lie algebras all continuous
subgroups of the Lorentz group were classified. It turned out that
the Lorentz group contains no 5-parameter subgroups and has only
one (up to isomorphism) 4-parameter subgroup. This subgroup
includes independent rotations around an arbitrarily selected
axis, the direction of which will be denoted using a unit
vector~$\boldsymbol\nu$, and \mbox{a~3-pa\-rameter} group
consisting of noncompact transformations only. Physically such
noncompact transformations are realized as follows. First choose
as $\boldsymbol\nu$ a direction towards a preselected star and
then perform an arbitrary Lorentz boost and complement it with
such a turn of the spatial axes that in a new reference frame the
direction towards the star remains unchanged.

The set of the transformations described, while linking the
inertial reference frames, actually constitutes a 3-parameter
noncompact group (in contrast to the usual Lorentz boosts). Let us
write the corresponding 3-parameter transformations in the
infinitesimal form
\begin{gather}
dx^0 = - \boldsymbol{n} \boldsymbol{x} d \alpha, \nonumber\\
d\boldsymbol{x} = (- \boldsymbol{n} x^0 -
[\boldsymbol{x}[\boldsymbol{\nu}\boldsymbol{n}]])d\alpha,\label{eq1}
\end{gather}
where $d\alpha$  is a rapidity, the unit vector $\boldsymbol{n}$
indicates a direction of the infinitesimal boost, so that
$d\boldsymbol{v} = \boldsymbol{n}d\alpha$, and the meaning of
$\boldsymbol{\nu}$ has been explained\footnote{For more details
see \cite{bogoslovsky1977}.}.
 Integration of equations \eqref{eq1} leads to the finite transformations
which, at any fixed $\boldsymbol\nu$, belong to the Lorentz group
and themselves form a 3-parameter noncompact group with parameters
${\boldsymbol n}$, $\alpha$:
\begin{gather}
\label{eq2} x^{'i}
=\Lambda^i_k(\boldsymbol{\nu};\boldsymbol{n},\alpha )\,x^k\,,
\end{gather}
where
\begin{gather*}
\Lambda^0_0=1+\frac{\cosh{\boldsymbol{\nu n}\alpha}-1}{(\boldsymbol{\nu n})^2},\\
\Lambda^0_{\beta}=\frac{1-e^{-\boldsymbol{\nu
n}\alpha}}{\boldsymbol{\nu n}}
\,n_{\beta}+\frac{\cosh{\boldsymbol{\nu n}\alpha}-1}{(\boldsymbol{\nu n})^2}\,\nu_{\beta},\\
\Lambda^{\rho}_0= \frac{1-e^{\boldsymbol{\nu
n}\alpha}}{\boldsymbol{\nu n}}\,n^{\rho}
+\frac{\cosh{\boldsymbol{\nu n}\alpha}-1}{(\boldsymbol{\nu n})^2}\,\nu^{\rho},\\
\Lambda^{\rho}_{\beta}=\delta^{\rho}_{\beta}+\frac{1-e^{\boldsymbol{\nu
n}\alpha}}{\boldsymbol{\nu n}}
\,n^{\rho}\nu_{\beta}+\nu^{\,\rho}\Lambda^0_{\beta}.
\end{gather*}
Hereafter the Latin indices take on values of 0, 1, 2, 3 while the
Greek ones, values of 1, 2, 3. Note also that the $n^{\,\beta}$
and ${\nu}^{\,\beta}$ denote the Cartesian components of unit
vectors  ${\boldsymbol n} $ and $\boldsymbol\nu$, in which case
$n_{\beta}=-n^{\beta}$, ${\nu}_{\beta}=-{\nu}^{\beta}$. The
transformations inverse to \eqref{eq2} appear as
\begin{gather}
\label{eq3} x^{i}
={\Lambda^{-1}}^i_k(\boldsymbol{\nu};\boldsymbol{n},\alpha)\,x'^{k},
\end{gather}
where
\begin{gather}
\label{eq4}
{\Lambda^{-1}}^i_k(\boldsymbol{\nu};\boldsymbol{n},\alpha
)=\Lambda^i_k (\boldsymbol{\nu};\boldsymbol{n},-\alpha).
\end{gather}

Consider two arbitrary elements of the group \eqref{eq2}. Let the
first element $g_1$ be characterized by the pa\-ra\-me\-ters
${\boldsymbol n}_1$,  $\alpha_1$, and the second one, $g_2$, by
the pa\-ra\-meters ${\boldsymbol n}_2$, $\alpha_2$. Then the
element   $g=g_2g_1$ will have the corresponding parameters
${\boldsymbol n}$, $\alpha$, which are functionally dependent on
${\boldsymbol n}_1$, $\alpha_1$ and  ${\boldsymbol n}_2$,
$\alpha_2$, i.e.\
$\Lambda^i_k(\boldsymbol{\nu};\boldsymbol{n},\alpha)=
\Lambda^i_j(\boldsymbol{\nu};{\boldsymbol n}_2,{\alpha}_2)
\Lambda^j_k(\boldsymbol{\nu};{\boldsymbol n}_1,{\alpha}_1)$. Using
the explicit form of the matrix elements $\Lambda_k^i(\boldsymbol
{\nu};{\boldsymbol n},\boldsymbol\alpha)$ and making the relevant
calculations, we arrive at the following relations:
\begin{gather}
\label{eq5} {\boldsymbol n}\,\alpha = \frac{\boldsymbol\nu
({\boldsymbol n}_1{\alpha}_1 +{\boldsymbol
n}_2{\alpha}_2)}{1-e^{\boldsymbol\nu ({\boldsymbol n}_1{\alpha}_1
+{\boldsymbol
n}_2{\alpha}_2)}}\left[\frac{1-e^{\boldsymbol\nu{\boldsymbol
n}_1{\alpha}_1}} {\boldsymbol\nu{\boldsymbol n}_1}\,{\boldsymbol
n}_1 + \frac{e^{\boldsymbol\nu{\boldsymbol n}_1{\alpha}_1}
(1-e^{\boldsymbol\nu{\boldsymbol n}_2{\alpha}_2})}{\boldsymbol\nu{\boldsymbol n}_2}\,{\boldsymbol n}_2\right],\\
\label{eq6} {\boldsymbol n}^2 = 1.
\end{gather}
These relations essentially represent the law of group composition
for the 3-parameter noncompact subgroup \eqref{eq2} of the Lorentz
group.

Since the group \eqref{eq2} links the coordinates of events in the
initial and primed inertial reference frames, from the physical
standpoint it is more natural to use as group parameters three
velocity components, $\boldsymbol v$, of the primed reference
frame rather than the $\boldsymbol{n}$, $\alpha$. In order to
express the $\boldsymbol v$ in terms of the $\boldsymbol{n}$,
$\alpha$ it is sufficient to consider in the initial frame, the
motion  of the origin of the primed frame, i.e.\ to write
$x^{\beta}={\Lambda^{-1}}^{\beta}_0 x'^{0}$ and
$x^{0}={\Lambda^{-1}}^0_0 x'^{0}$. Then
$v^{\beta}=x^{\beta}/x^{0}={\Lambda^{-1}}^{\beta}_0/
{\Lambda^{-1}}^0_0$. Using equations~\eqref{eq2}--\eqref{eq4}, we
get as a result
\begin{gather}
\label{eq7} \boldsymbol v=\left[\frac{1-e^{-\boldsymbol{\nu
n}\alpha}}{\boldsymbol{\nu n}}\, \boldsymbol n
+\frac{\cosh{\boldsymbol{\nu n}\alpha}-1}{(\boldsymbol{\nu
n})^2}\, \boldsymbol\nu
\right]\biggl/\left[1+\frac{\cosh{\boldsymbol{\nu n}\alpha}-1}
{(\boldsymbol{\nu n})^2}\right].
\end{gather}
Hereafter we put  $c=1$. Since ${\boldsymbol n}^2 =
{\boldsymbol\nu}^2 = 1\,$, equation~\eqref{eq7} yields the inverse
relations:
\begin{gather}
\label{eq8} \boldsymbol n =\frac{\boldsymbol
v}{\sqrt{2(1-\boldsymbol{v\nu })(1-\sqrt{1- {\boldsymbol v}^2})}}
- \sqrt{\frac{1-\sqrt{1-{\boldsymbol v}^2}} {2(1-\boldsymbol{v\nu
})}}\,\boldsymbol{\nu},
\\
\label{eq9} \alpha =\frac{\sqrt{2(1-\boldsymbol{v\nu
})(1-\sqrt{1-{\boldsymbol v}^2})}} {\sqrt{1-{\boldsymbol
v}^2}+\boldsymbol{v\nu }-1}\, \ln\left(\frac{\sqrt{1-{\boldsymbol
v}^2}}{1-\boldsymbol{v\nu }}\right).
\end{gather}
In terms of $\boldsymbol v$ the law of group composition
\eqref{eq5}, \eqref{eq6} takes the form
\begin{gather}
\label{eq10} \boldsymbol v =\frac{\big({\boldsymbol v}_1(1-
{\boldsymbol v}_2\boldsymbol\nu )+{\boldsymbol
v}_2\sqrt{1-{\boldsymbol v}_1^2} \,\big)(1-{\boldsymbol v}_1
\boldsymbol\nu )+\boldsymbol\nu \big({\boldsymbol v}_1
{\boldsymbol v}_2+\boldsymbol\nu{\boldsymbol v}_2 \big(\sqrt{1-
{\boldsymbol v}_1^2}-1\big)\big)\sqrt{1-{\boldsymbol
v}_1^2}}{1-{\boldsymbol v}_1 \boldsymbol\nu +{\boldsymbol
v}_1{\boldsymbol v}_2\sqrt{1-{\boldsymbol v}_1^2}+
\boldsymbol\nu{\boldsymbol v}_2\big(1-{\boldsymbol
v}_1\boldsymbol\nu +\sqrt{1- {\boldsymbol
v}_1^2}\,\big)\big(\sqrt{1-{\boldsymbol v}_1^2}-1\big)}, \!\!\!
\end{gather}
whereby $\Lambda^i_k(\boldsymbol{\nu};\boldsymbol{v})=
\Lambda^i_j(\boldsymbol{\nu};{\boldsymbol v}_2)
\Lambda^j_k(\boldsymbol{\nu};{\boldsymbol v}_1)$. It is thus clear
that equation~\eqref{eq10} represents the Einstein law of
addition of velocities  ${\boldsymbol v}_1$ and  ${\boldsymbol
v}_2$. In comparison with its usual form one should remember,
however, that after the transformation
$\Lambda^j_k(\boldsymbol{\nu};{\boldsymbol v}_1)$ the spatial
axes, in which the ${\boldsymbol v}_2$ is prescribed, appear now
to be not parallel to the initial axes, but turned so that the
vector $\boldsymbol\nu$ relative to them maintains its initial
orientation. Therefore, it is true, irrespective of the direction
of ${\boldsymbol v}_1$, equation~\eqref{eq10} yields $\boldsymbol
v = \boldsymbol\nu$ if  ${\boldsymbol v}_2 = {\boldsymbol\nu}$.

\subsection{3-parameter group of the generalized Lorentz transformations}

The 3-parameter group of the generalized Lorentz transformations
(of the generalized Lorentz boosts), similarly to the subgroup
\eqref{eq2} of Lorentz group, consists of noncompact
transformations only. In the infinitesimal form the
transformations belonging to it appear as
\begin{gather}
dx^0 = (-r(\boldsymbol{\nu}\boldsymbol{n})x^0 - \boldsymbol{n} \boldsymbol{x}) d \alpha, \nonumber\\
d\boldsymbol{x} =
(-r(\boldsymbol{\nu}\boldsymbol{n})\boldsymbol{x} -
 \boldsymbol{n} x^0 - [\boldsymbol{x}[\boldsymbol{\nu}\boldsymbol{n}]])d\alpha.
\label{eq11}
\end{gather}
Here, as in the infinitesimal transformations \eqref{eq1} of the
group \eqref{eq2}, the ${\boldsymbol n}$ and  $\alpha$ are group
parameters, while the $\boldsymbol\nu$ is a fixed unit vector. And
the difference between \eqref{eq11} and \eqref{eq1} consists in
the appearance of an additional generator of the scale
transformations, which is proportional to a new fixed
dimensionless parameter $r$. Since the scale transformations
commute with the Lorentz boosts and 3D rotations, the result of
integration of equations \eqref{eq11} is a priori clear:
\begin{gather}
\label{eq12} x'^{i}
=D(r,\boldsymbol{\nu};\boldsymbol{n},\alpha)\Lambda^i_k(\boldsymbol{\nu};\boldsymbol{n},\alpha)x^k,
\end{gather}
where
$D(r,\boldsymbol{\nu};\boldsymbol{n},\alpha)=e^{-r{\boldsymbol\nu}{\boldsymbol
n}\alpha}I$,
 whereby  $I$ is a unit matrix while $\Lambda^i_k(\boldsymbol{\nu};\boldsymbol{n},\alpha )$
 are matrices which make up the group~\eqref{eq2}. As for the law
 of group composition for the group \eqref{eq12}, it is given by equations~\eqref{eq5}
 and \eqref{eq6} obtained for the group \eqref{eq2}. We note incidentally that equation~\eqref{eq5}
 yields the relation
$\boldsymbol\nu{\boldsymbol n}\alpha = \boldsymbol\nu{\boldsymbol
n}_1{\alpha}_1 + \boldsymbol\nu{\boldsymbol n}_2{\alpha}_2$. Thus
the group of the generalized Lorentz transformations~\eqref{eq12},
on the one hand, is locally isomorphic to the corresponding
3-parameter subgroup \eqref{eq2} of the Lorentz group and, on the
other hand, it is a homogeneous 3-parameter noncompact subgroup of
the similitude group \cite{patera&winternitzzassenhaus}. Since
(according to \eqref{eq12}) in passing to the primed inertial
frame the time $(x^0)$ and space $(\boldsymbol x)$ event
coordinates undergo identical additional dilatations~$D$, then the
velocity $\boldsymbol v$ of the primed frame is related to the
group parameters ${\boldsymbol n}$, $\alpha$ by the same
equations~\eqref{eq7}--\eqref{eq9} as in the case $r=0$ where the
group \eqref{eq12} coincides with the subgroup \eqref{eq2} of the
Lorentz group. For the same reason the transformations
\eqref{eq12} retain valid Einstein's law of addition of
3-velocities, written as \eqref{eq10}. As for the
reparametrization of the group \eqref{eq12}, then, for example,
the matrix $D$, involved in \eqref{eq12}, takes the following form
in terms of $\boldsymbol v$:
\begin{gather}
\label{eq13} D(r,\boldsymbol\nu;\boldsymbol v
)=\left(\frac{1-\boldsymbol v\boldsymbol\nu} {\sqrt{1-\boldsymbol
v^{\,2}}} \right)^r I.
\end{gather}

Unlike the transformations \eqref{eq2}, the 3-parameter group of
the generalized Lorentz boosts \eqref{eq12} conformally modifies
the Minkowski metric but leaves invariant the metric
\begin{gather}\label{eq14}
ds^2=\left[\frac{(dx_0-\boldsymbol\nu d\boldsymbol
x)^2}{dx_0^2-d\boldsymbol x^{2}}\right]^r \big(dx_0^2-d\boldsymbol
x^{2}\big).
\end{gather}
The given Finslerian metric generalizes the Minkowski metric and
describes the relativistically invariant flat space of events with
partially broken 3D isotropy. The inhomogeneous isometry group of
space \eqref{eq14} is an 8-parameter one: four parameters
correspond to space-time translations, one parameter, to rotations
about the physically preferred direction $\boldsymbol\nu$, and
three parameters, to the generalized Lorentz boosts.

\subsection{Bispinor representation of the 3-parameter group\\ of generalized Lorentz boosts}

Now turn to the construction of bispinor representation of the
group of the generalized Lorentz boosts \eqref{eq12}. Since the
group \eqref{eq12} is locally isomorphic to the 3-parameter
subgroup \eqref{eq2} of the Lorentz group, its bispinor
representation
 must also be locally isomorphic to the bispinor representation
 of the subgroup \eqref{eq2}. This means that the
 transformations $x'^i = D \Lambda_k^i x^k$ of
 the event coordinates should be accompanied by the following transformations of a bispinor field:
$\Psi'(x') = D^dS\Psi(x)$, ${\overline{\Psi}\,}'(x') =
\overline{\Psi}(x)D^dS^{-1}$. Here the group of matrices $S$,
operating on the bispinor indices, represents the subgroup
\eqref{eq2} of the Lorentz matrices $\Lambda_k^i$ while  $D^d$
denotes the corresponding additional scale transformations of
bispinors, in which case the unit matrix, involved in the
definition \eqref{eq13}, operates on the bispinor indices in this
context. Since $d^4x' =
 |D\Lambda_k^i|d^4x = D^4d^4x$ and matrices $S$ satisfy the standard
 condition $S^{-1}\gamma^n S = \Lambda_m^n\gamma^m$, then, proceeding
 from the generalized Lorentz invariance of action for a free
 massless field $\Psi$, it is easy to show that $d = -3/2$. As
 a result the bispinor representation of the group of generalized Lorentz
 boosts \eqref{eq12} is realized by the transformations
\begin{gather}
\label{eq15} \Psi'= D^{-3/2}S\Psi, \qquad
{\overline{\Psi}\,}'=\overline{\Psi}D^{-3/2}S^{-1}
\end{gather}
and it remains to find a 3-parameter group of the matrices $S =
S(\boldsymbol\nu;\boldsymbol n,\alpha)$. For this purpose, using
the 4-vectors $\nu^l=(1, \boldsymbol{\nu})$; $\nu_l=(1,
-\boldsymbol{\nu})$; $n^l=(0, \boldsymbol{n})$; $n_l=(0,
-\boldsymbol{n})$, first rewrite the infinitesimal transformations
\eqref{eq1} in the form $dx^i = {\omega^i}_k x^k$, where
${\omega^i}_k =(\nu^in_k - n^i \nu_k) d \alpha$, in which case
$-\omega_{ki} = \omega_{ik} = (\nu_in_k - \nu_kn_i)d\alpha$. Thus,
in the vicinity of the identical transformation the matrices
$\Lambda_k^i$ take the form $\Lambda_k^i(d\alpha)=\delta^i_k +
{\omega^i}_k$. Respectively, the $S(d\alpha)=1+\frac{1}{8}
(\gamma^i \gamma^k - \gamma^k \gamma^i)\omega_{ik}$. Considering
that $n_0=0$ and $\nu_0 = 1$, the latter relation leads to $S$ in
the form:
\begin{gather}
\label{eq16} S(\boldsymbol\nu;\boldsymbol n,\alpha) = e^{\left\{
\cdots \right\}{\alpha}/2}.
\end{gather}
Here and below we use the notation $\left\{ \cdots \right\}$ for
the sum of generators of Lorentz boosts and 3-rotations about the
vector $[\boldsymbol\nu \boldsymbol n]$, that is
\begin{gather}
\label{eq17} \left\{ \cdots \right\} =
-\gamma^0\boldsymbol\gamma\boldsymbol{n} -
i\boldsymbol{\Sigma}[\boldsymbol{\nu} \boldsymbol{n}],
\end{gather}
where ${\gamma}^0$, $\boldsymbol\gamma$ are the Dirac matrices,
$\boldsymbol\Sigma ={\rm diag}\,(\boldsymbol\sigma,
\boldsymbol\sigma)$ and $\boldsymbol \sigma$ are the Pauli
matrices. With the aid of the algebra of $\gamma$ matrices one can
find that
\begin{gather}\nonumber
\begin{array}{lll}
\left\{ \cdots \right\}^2 = ({\boldsymbol \nu}{\boldsymbol n})^2
I, & \left\{ \cdots \right\}^4
=({\boldsymbol \nu}{\boldsymbol n})^4 I, & \ldots ;\vspace{1mm}\\
\left\{ \cdots \right\}^3 = ({\boldsymbol \nu}{\boldsymbol
n})^3\left\{
  \cdots \right\}/({\boldsymbol \nu}{\boldsymbol n}),\quad &
\left\{ \cdots \right\}^5 = ({\boldsymbol \nu}{\boldsymbol
n})^5\left\{
  \cdots \right\}/({\boldsymbol \nu}{\boldsymbol n}),\quad & \ldots .\\
\end{array}
\end{gather}
These relations make it possible to represent \eqref{eq16} as
\begin{gather}
\label{eq18} S(\boldsymbol\nu;\boldsymbol n,\alpha) =I\cosh
\frac{{\boldsymbol \nu}{\boldsymbol n}\alpha}{2} -
 \frac{i\boldsymbol{\Sigma}[\boldsymbol{\nu} \boldsymbol{n}]
 +\gamma^0\boldsymbol\gamma\boldsymbol{n}}{{\boldsymbol \nu}{\boldsymbol n}}
\sinh \frac{{\boldsymbol \nu}{\boldsymbol n}\alpha}{2}.
\end{gather}
Now reparametrizing the $S(\boldsymbol \nu;\boldsymbol n,\alpha)$
with the aid of \eqref{eq8}, \eqref{eq9} and using equations
\eqref{eq15} and \eqref{eq13}, we thus arrive at the following
3-parameter noncompact group of bispinor transformations in the
axially symmetric flat Finslerian space~\eqref{eq14}:
\begin{gather}
\Psi'=\frac{\left ((1-\boldsymbol v\boldsymbol\nu
)/\sqrt{1-{\boldsymbol v}^2} \right
)^{-3r/2}}{2\sqrt{(1-\boldsymbol v\boldsymbol\nu
)\sqrt{1-{\boldsymbol v}^2}}}
\left\{\left (1-\boldsymbol v\boldsymbol\nu + \sqrt{1-{\boldsymbol v}^2}\right )I\right.  \nonumber \\
\phantom{\Psi'=}{}- \left. i\,[\boldsymbol{\nu}
\boldsymbol{v}]\boldsymbol{\Sigma} -
\left(\boldsymbol{v}-(1-\sqrt{1-{\boldsymbol
v}^2})\,\boldsymbol\nu \right ) \gamma^0\boldsymbol\gamma \right\}
\Psi .\label{eq19}
\end{gather}
We note finally that an invariant of the transformations
\eqref{eq19} is the Finslerian form
\begin{gather*}
[({\nu_n
\overline{\Psi}\gamma^n\Psi}/{\overline{\Psi}\Psi})^2]^{-3r/2}\overline{\Psi}\Psi
\end{gather*}
but no longer the bilinear form $\overline \Psi \Psi$.

\section{Finite transformations belonging to two  subgroups\\
   of the 3-parameter group of  generalized Lorentz boosts}

\subsection{Abelian 2-parameter subgroup and its geometric invariants}

The Abelian 2-parameter subgroup is generated by the
infinitesimal  transformations \eqref{eq11} provided
 that $\boldsymbol{n} \perp \boldsymbol{{\nu}}$, and its finite transformations can
 be derived from the finite 3-parameter transformations \eqref{eq12}
 via the passage to the limit $(\boldsymbol{\nu n} \alpha ) \rightarrow 0$. As a result, we find
\begin{gather}
 x'_0  =  \left( \frac{\alpha^2}{2} + 1 \right)
 x_0 - \alpha (\boldsymbol{nx}) - \frac{\alpha^2}{2} (\boldsymbol{\nu x}),
 \nonumber \\
 {\boldsymbol x}'  =  {\boldsymbol x} + {\boldsymbol n} (- x_0 + \boldsymbol{\nu x}) \alpha +
 \boldsymbol{\nu} \left[
 (x_0 - \boldsymbol{\nu x})  \frac{\alpha^2}{2} -
 (\boldsymbol{nx}) \alpha \right]. \label{eq20}
 \end{gather}
 The inverse transformations are obtained by means of
 the substitution $\alpha \rightarrow -\alpha$:
\begin{gather}
 x_0  =  \left( \frac{\alpha^2}{2} + 1\right)x'_0 +
 \alpha (\boldsymbol{nx}') - \frac{\alpha^2}{2} (\boldsymbol{\nu x}'), \nonumber\\
 {\boldsymbol x}  =  {\boldsymbol x}' + {\boldsymbol n} ( x'_0 - \boldsymbol{\nu x}') \alpha +
 \boldsymbol{\nu} \left[ (x'_0 - \boldsymbol{\nu x}')  \frac{\alpha^2}{2} +
 (\boldsymbol{nx}') \alpha \right]. \label{eq21}
\end{gather}
As before, in order to express the velocity $\boldsymbol v$ of the
primed inertial frame  in terms of the parame\-ters $\boldsymbol
n$, $\alpha$ it is sufficient to consider in the initial frame,
the motion of the origin of the primed frame, i.e.\ to  put
$\boldsymbol x' = 0$ . Then we get $x_0=(\alpha^2/2+1)x'_0$,
$\boldsymbol x=(\boldsymbol n\alpha +
\boldsymbol{\nu}\alpha^2/2)x'_0$. As a result,
\begin{gather}\label{eq22}
 \boldsymbol v =
 \frac{\boldsymbol n \alpha + \boldsymbol{\nu} \alpha^2/2}{1+
  \alpha^2/2 }.
\end{gather}
It is easy to verify that in terms of $\boldsymbol v$ the
condition $\boldsymbol{n} \perp \boldsymbol{{\nu}}$ takes the form
\begin{gather}\nonumber
 \left( \frac{1- \boldsymbol{v\nu}}{\sqrt{1-\boldsymbol v^2}} \right)^2  =  1.
 \end{gather}
Thus, having used the equation \eqref{eq22} for reparametrization
of the initial transformations  \eqref{eq20}, we get
\begin{gather}
 x'_0  =  \frac{x_0 - \boldsymbol{vx}}{1-\boldsymbol{v\nu}},
 \nonumber\\
 \boldsymbol{x}'  =   \boldsymbol{x} -
 \frac{(x_0 - \boldsymbol{\nu x})\boldsymbol v -
 \left[ (2x_0 - \boldsymbol{\nu x}) (\boldsymbol{v\nu}) -
 (\boldsymbol{vx})\right]\boldsymbol\nu}{1- \boldsymbol{v\nu}}. \label{eq23}
 \end{gather}
One can verify that these transformations have two independent
invariants:
 $(x_0^2 - \boldsymbol x^2)$ and $(x_0 - \boldsymbol{\nu x})$.
Therefore the transformations \eqref{eq23} belong to the
6-parameter Lorentz group. At the same time, leaving the
Finslerian metric \eqref{eq14} invariant, they make up the
2-parameter Abelian subgroup of the 3-parameter group of
generalized Lorentz boosts \eqref{eq12}.

In spite of a new ( Finslerian ) geometry of the event space the
3-velocity space remains the Lobachevski space. This means that
the generalized Lorentz boosts \eqref{eq12} induce the
corresponding motions of the Lobachevski space. In particular, the
above-described Abelian 2-parameter  boosts \eqref{eq23} induce
such motions which leave invariant a family of the horospheres
\begin{gather*}\nonumber
 \frac{1- \boldsymbol{v\nu}}{\sqrt{1-\boldsymbol v^2}} =
{\rm  const},
 \end{gather*}
i.e.\ of surfaces perpendicular to the congruence of geodesics
parallel to $\boldsymbol{\nu}$ and possessing Euclidean inner
geometry.

\subsection{1-parameter subgroup and its geometric invariants}

Finite transformations, making up the 1-parameter subgroup of the
3-parameter group of gene\-ralized Lorentz boosts, follow from the
3-parameter finite transformations \eqref{eq12} provided that
${\boldsymbol{n}}
 \parallel {\boldsymbol{\nu}}$. As a result we get
\begin{gather}
 x'_0  =  e^{-r\alpha} \left\{ x_0\cosh\alpha - (\boldsymbol{\nu x}) \sinh\alpha
 \right\}, \nonumber \\
 \boldsymbol x'  =  e^{-r\alpha} \left\{ \boldsymbol x - \boldsymbol{\nu} (\boldsymbol{\nu x}) +
 \boldsymbol{\nu}
 \left[ -x_0\sinh\alpha + (\boldsymbol{\nu x})\cosh\alpha \right]  \right\}.\label{eq24}
 \end{gather}
The inverse 1-parameter transformations have the form
\begin{gather}
 x_0  =  e^{r\alpha} \left\{ x'_0\cosh\alpha + (\boldsymbol{\nu x}')\sinh\alpha
 \right\} ,  \nonumber \\
 \boldsymbol x  =  e^{r\alpha} \left\{ \boldsymbol x'- \boldsymbol{\nu} (\boldsymbol{\nu x}') +
 \boldsymbol{\nu}
 \left[ x'_0\sinh\alpha + (\boldsymbol{\nu x}')\cosh\alpha  \right] \right\} . \label{eq25}
\end{gather}
According to equations~\eqref{eq25}, the velocity $\boldsymbol v$
of the primed system is related
 to the $\boldsymbol\nu $ and the group parameter $\alpha$  by the relations
\begin{gather*}
\boldsymbol v  =  \boldsymbol{\nu} \tanh\alpha ,
\\
\frac{\boldsymbol{v\nu}}{\sqrt{1-{\boldsymbol v}^2}} =
\sinh\alpha,\qquad
 \frac{1}{\sqrt{1-{\boldsymbol v}^2}} = \cosh\alpha,\qquad
 \frac{1- \boldsymbol{v\nu}}{\sqrt{1-\boldsymbol v^2}}=e^{-\alpha}.
\end{gather*}
Since
\begin{gather*}\nonumber
(x'_0 - \boldsymbol{\nu x}') = e^{(1-r)\alpha} (x_0 -
\boldsymbol{\nu x}),
\\
({x'_0}^2 - {\boldsymbol x'}^2) = e^{-2r\alpha} (x_0^2 -
\boldsymbol x^2),
\\
[\boldsymbol x'\boldsymbol\nu ]  =  e^{-r \alpha}
[\boldsymbol{x\nu }],
\end{gather*}
we see that the 1-parameter  group of  generalized Lorentz boosts
\eqref{eq25} conformally modifies the Minkowski
 metric, but leaves invariant the Finslerian metric \eqref{eq14}.
In addition, we get the following invariant of this group:
\begin{gather*}\nonumber
\frac{[\boldsymbol{x\nu }]} {\sqrt{x^2_0 - \boldsymbol x^2}} =
{\rm invar}.
\end{gather*}

In the Lobachevski space of the 3-velocities  $\boldsymbol v$ the
1-parameter transformations under consi\-de\-ration induce such
motions which leave invariant a family of the surfaces
\begin{gather*}\nonumber
 \frac{\boldsymbol v^2 - (\boldsymbol{v\nu })^2}{1 - \boldsymbol v^2}=
 {\rm const}.
\end{gather*}
These surfaces have the form of cylinders formed by the rotation
of
 the equidistant lines about~$\boldsymbol{\nu}$.

\section{Conclusion}

Within the framework of the Finslerian approach to the problem of
violation of the Lorentz symmetry we have studied the flat axially
symmetric Finslerian space of events which is a gene\-ralization of
Minkowski space. Such event space arises from spontaneous breaking
of initial gauge symmetry and from formation of an anisotropic
fermion-antifermion condensate. We have shown that the appearance
of an anisotropic condensate breaks the Lorentz symmet\-ry but the
relativistic symmetry, realized by means of the 3-parameter group
of generalized Lorentz boosts, remains valid here. The principle
of generalized Lorentz invariance enables exact taking into
account the influence of an axially symmetric condensate on the
dynamics of fundamental fields. In particular, with its aid in
\cite{bogoslovsky&goenner200104}, the Lagrangian, describing the
dynamics of the massive fermion field in the relativistically
invariant anisotropic condensate, has been constructed. This
Lagrangian leads to some nonlinear spinor equations. The
algebraic-theoretical methods of constructing exact solutions for
a wide class of nonlinear spinor equations, developed by
W.I.~Fushchych with coworkers \cite{fushchych&zhdanov}, make it
possible to  find the so-called nongenerable solutions of the
above-mentioned spinor equations. Just with this objective in view
the subgroups of the group of generalized Lorentz transformations
and their geometric invariants have been studied in the present
work.

\subsection*{Acknowledgements}

The author is grateful to Prof.~H.~Goenner for the fruitful
collaboration that led to the results presented in this paper. It
is pleasure to thank also  Prof.~R.~Tavakol for helpful
discussions.

\LastPageEnding

\end{document}